\documentclass[12pt]{article}
\usepackage[all]{xy}

\usepackage{epic}
\usepackage{amsmath}[1996/11/01]
\usepackage{amssymb,amsthm,amsfonts,latexsym,epsfig,graphics}
 \def\beql#1#2\eeql{\begin{equation}\label{#1}#2\end{equation}}

\DeclareMathOperator{\Gal}{Gal}

\DeclareMathOperator{\rk}{rk}

\DeclareMathOperator{\End}{End}
\DeclareMathOperator{\GL}{GL}
\DeclareMathOperator{\Aut}{Aut}

\newtheorem{theorem}{Theorem}[section]

\newtheorem{cor}[theorem]{Corollary}

\newcommand{\bew}{\noindent\underline{Proof.}\ }
\newtheorem{rem}[theorem]{Remark}
\newtheorem{lemma}[theorem]{Lemma}
\newtheorem{proposition}[theorem]{Proposition}
\newtheorem{definition}[theorem]{Definition}

\newcommand{\disj}{\stackrel{.}{\cup}}

\newcommand{\N}{{\mathbb{N}}}

\newcommand{\cC}{{\mathcal{C}}}
\newcommand{\cD}{{\mathcal{D}}}
\newcommand{\cG}{{\mathcal{G}}}
\newcommand{\cM}{{\mathcal{M}}}
\newcommand{\MRD}{\rm MRD\ }
\newcommand{\eb}{\phantom{zzz}\hfill{$\square $}\smallskip}

\begin{document}
\begin{center}
{\Large {\bf Automorphism groups of Gabidulin-like codes}} \\
\vspace{1.5\baselineskip}
{Dirk Liebhold and Gabriele Nebe\footnote{
Lehrstuhl D f\"ur Mathematik, RWTH Aachen University,
52056 Aachen, Germany, 
 nebe@math.rwth-aachen.de} }
\end{center}

{\sc Abstract.}
{\small 
Let $K/k$ be a cyclic Galois extension of degree $\ell $ and $\theta $ a generator of $\Gal(K/k)$. 
For any $v=(v_1,\ldots , v_m)\in K^m$ such that $v$ is linearly independent over $k$, 
and any $1\leq d < m $ 
the Gabidulin-like code $\cC(v,\theta , d) \leq k^{\ell \times m }$ is a maximum
rank distance code of dimension $\ell d$ over $k$. 
This construction unifies the ones available in the literature. 
We characterise the $K$-linear codes that are Gabidulin-like codes and 
determine their rank-metric automorphism group.
\\[1ex]
 \MRD codes,  automorphism group,  Gabidulin-like code
\\
}

\section{Introduction.}

In random linear network coding any node of the network may 
transmit a random linear combination of the received vectors.
So the transmitted information is the subspace
generated the input vectors, an element of the Grassmannian 
$$\cG _{\ell,n}(k) := \{ U \leq k^n \mid \dim (U) = \ell \} $$
the set of all $\ell $-dimensional subspaces of the space $k^n$ of
rows of length $n$ over the field $k$. 
A (constant dimension) network code is a subset of such a Grassmannian. 
There is a natural distance function $d$ on $\cG _{\ell,n}(k)$ 
defined by $d(U,V) := \ell - \dim(U\cap V) $. 
The general linear group $\GL_n(k)$ acts transitively on $\cG _{\ell,n}(k)$ 
preserving this distance.
However there are a few disadvantages of this framework: 
\begin{itemize}
\item $\cG _{\ell ,n}(k)$ is a homogeneous space but not a vector space.
\item So in this generality there is no notion of a linear code (as for the
classical block codes). 
\item It is also not obvious how to systematically 
encode information into a sequence of subspaces. 
\end{itemize} 
To come around these problems, Koetter and Kschischang \cite{KK} suggested 
to consider a subset of $\cG_{\ell,n}(k) $:  Put $m:=n-\ell $.
For a matrix $X\in k^{\ell\times m} $ let 
$U_X := $ row space of $(I_{\ell} | X)  $.
Then $U_X  \in \cG _{\ell,n}(k)$ and $U_X=U_Y$ if and only if $X=Y$.
So the map $X\mapsto U_X$ is a bijection between the 
vector space $k^{\ell \times m }$ and 
$$\cM _{\ell,m}(k) := \{ U_X \mid X \in k^{\ell\times m}  \} 
\subset \cG _{\ell,n}(k).$$
The distance between two spaces $U_X,U_Y \in \cM _{\ell,n}(k)  $ 
is $d(U_X,U_Y) = \rk (X-Y)$, the rank metric on this space of matrices, 
which is
studied in \cite{Delsarte} and \cite{Gabidulin}.

A linear rank metric code is a subspace $\cC $ of $k^{\ell\times m} $. 
The minimum distance of $\cC $ is $d(\cC )= \min \{ \rk (C) \mid 0 \neq C \in \cC \}$. 
The well known Singleton bound (see Proposition \ref{singleton}) 
shows that $$\dim(\cC ) \leq \max (\ell ,m) (\min(\ell , m) - d(\cC) +1) .$$
Codes where equality holds are called maximum rank distance 
(or \MRD) codes.

%The $k$-linear rank distance preserving automorphisms of  $k^{\ell\times m} $ 
%are the 
%$$\kappa _{g,h} : X \mapsto g^{-1} X h \mbox{ for } g\in \GL_\ell (k) , h\in \GL _m (k) $$
%(see \cite[Theorem 3.4]{Wan}) and, if $\ell = m$, also 
%$$X \mapsto g^{-1} X^{tr} h  \ \ ( g\in \GL_\ell (k) , h\in \GL _m (k) ) .$$
%Two codes $\cC $ and $\cD $ in $k^{\ell\times m} $ are called 
%(properly) equivalent, if $\cC = g^{-1} \cD h $ for some $g\in \GL_\ell (k) , h\in \GL _m (k)$
%and 
%$$\Aut (\cC ) := \{ (g,h) \in  \GL_\ell (k) \times \GL _m (k) \mid 
%g^{-1} \cC h = \cC \} $$ is called the (proper) automorphism group of $\cC $. 

The most famous construction of \MRD codes is due to Gabidulin \cite{Gabidulin}. 
In this paper we define Gabidulin-like codes (Definition \ref{Gabi}) 
which provide a unified framework of various generalisations 
of Gabidulin codes.  Their basic properties are studied in 
Section \ref{MRD}, where we show that Gabidulin-like codes are \MRD codes and 
provide a characterisation (as in \cite{Trautmann}) 
which lifted codes are Gabidulin-like codes (Theorem \ref{cond}). 
Section \ref{AUTO} then 
describes an algorithm
to compute the automorphism group of rank metric codes
which can also be used to test equivalence. 
Using the strategy of this algorithm we will describe the automorphism 
groups of Gabidulin-like codes in Section \ref{GABI}.
In the special case of classical Gabidulin codes of full length $m=\ell$
these groups have already been determined in \cite{Willems} and \cite{Sheekey}.

\section{Rank metric codes.} \label{MRD} 

Let $k$ be any field, $\ell , m \in \N $. 
To simplify notation we will always assume that $\ell \geq m > 0$. 

\begin{definition}
A linear rank metric code is a subspace $\cC $ of $k^{\ell\times m} $. 
The minimum distance of $\cC $ is $d(\cC )= \min \{ \rk (C) \mid 0 \neq C \in \cC \}$. 
\end{definition}

The following analogue of the classical Singleton bound is well 
known for rank metric codes (\cite[Theorem 5.4]{Delsarte}, \cite[Lemma 1]{Gow}). 

\begin{proposition}\label{singleton} 
Let $\cC \leq k^{\ell \times m}$ be a rank metric code of 
dimension $d$ and minimum distance $r$. 
Then $d\leq \ell ( m - r +1 ).$
Codes that achieve equality are called 
\MRD codes (maximum rank distance codes). 
\end{proposition}

\bew
Let $\pi $ denote the projection of $k^{\ell \times m}$ onto 
$k^{\ell \times (m-r+1)}$ omitting the last $r-1$ columns of any matrix. 
Then clearly the kernel of this projection consists of matrices of 
rank $\leq r-1$. In particular the restriction of $\pi $ to 
$\cC $ is an injective mapping of $\cC $ into a space of
dimension $\ell (m-r+1) $ thus
$$d=\dim (\cC ) = \dim (\pi (\cC )) \leq \ell (m-r+1 ).$$
\eb

Clearly the dimension of a  maximum rank distance code is always a
multiple of $\ell $ but apart from this obvious restriction, 
\MRD codes exist for all possible parameters, if $k$  admits
a cyclic field extension $K$ of degree $\ell = [K:k]$
(see \cite[Lemma 3.2]{Guralnick} or Definition \ref{Gabi} below).
These examples have the property that they are linear over the
larger field. 
Until recently, \cite{Sheekey}, all known families of \MRD codes 
arose from linear codes over some extension field $K$, so called
lifted codes:

\begin{definition}
Let $K/k$ be a field extension of degree $\ell $ 
and $\tilde{\cC} \leq K^m$ a $K$-linear code of length $m$.
Fix some basis  $B = (B_1,\ldots , B_{\ell })  \in K^{\ell }$.
Then
$$\epsilon _B: K \to k^{\ell \times 1}, \epsilon _B (\sum _{i=1}^\ell a_i B_i ) = (a_1,\ldots , a_{\ell }) ^{tr} $$ 
maps  $\tilde{\cC }$ to 
the {lifted code} 
$$\cC := \epsilon _B (\tilde{\cC } )= \{ 
(\epsilon _B(c_1),\ldots , \epsilon _B(c_m ))  \mid 
(c_1,\ldots , c_m) \in \tilde{\cC } \} \leq k^{\ell \times m} .$$
\end{definition}

The lifted codes (with respect to the chosen $k$-basis $B$ of $K$) are exactly 
the codes $\cC \leq k^{\ell \times m}$ 
that are invariant under left multiplication 
with $\Delta _B(K)\leq k^{\ell \times \ell}$, the regular representation of $K$
with respect to $B$.

\begin{rem}\label{rank}
The rank of $\epsilon _B((v_1,\ldots , v_m)) $ equals the 
$k$-dimension of the subspace $\langle v_1,\ldots , v_m \rangle _k$ 
of $K$.
Therefore we call this dimension also the rank of the vector 
$v=(v_1,\ldots , v_m)\in K^m$.
\end{rem}

The most well known construction of an \MRD code as a lifted code 
is due to Gabidulin \cite{Gabidulin} (cf. \cite{KGabidulin} for a 
generalisation for finite fields and \cite{Augot} for a generalisation
to characteristic 0). All these constructions only depend on the
fact that $K/k$ is a cyclic Galois extension:

\begin{definition}\label{Gabi} 
Let $K/k$ be a cyclic field extension of 
degree $\ell $ and $\theta $ a generator of $ \Gal (K/k )$. 
For $v=(v_1,\ldots , v_m) \in K^m$ and 
 any $1\leq d\leq m-1$  we put 
$$\tilde{\cC }(v,\theta,d)  := \langle v,\theta (v) ,\ldots , \theta ^{d-1}(v) \rangle _K
\leq K^m $$
where $\theta ^j (v) = (\theta ^j(v_1),\ldots , \theta ^j(v_m)) $ 
and
$$\cC (v,\theta ,d) := \epsilon _B(\tilde{\cC }(v,\theta,d)) \leq k^{\ell \times m} .$$
If the rank of 
$v$  equals $m$ then $v$ is called a  Gabidulin vector and 
$\cC (v,\theta ,d) $ the  Gabidulin-like code with 
parameters $(v,\theta,d)$.
\end{definition}

It can easily be seen (see the proof of Theorem \ref{cond} below) that 
$\cC (v,\theta ,d) $ is not an \MRD code, if $v$ is not a 
Gabidulin vector.

\begin{lemma}\label{jordan} (cf. \cite[Theorem 1]{Augot},
 \cite[Lemma 3.2]{Guralnick})
Assume that $\Gal (K/k) = \langle \theta \rangle $ and
let $p = \sum _{i=0}^t p_i x^i \in K[x]$ be a non-zero polynomial of degree $t$. 
Then the kernel of 
$$p(\theta ):= \sum _{i=0}^t p_i \theta ^i \in  \End_k(K),\ \alpha 
\mapsto \sum _{i=0}^t p_i \theta ^i(\alpha ) $$
is a $k$-subspace of $K$ of dimension at most $t$.
\end{lemma} 

\bew
As $\theta $ is a generator of the Galois group of $K/k$ the 
powers $(1,\theta,\ldots , \theta ^{\ell -1} ) \in \End_k(K)^{\ell } $ 
are linearly independent over $K$ (see for instance the proof of Theorem (29.12) in \cite{Reiner}) and 
$$k^{\ell \times \ell }  \cong \End_k(K) \cong 
\bigoplus _{i=0}^{\ell -1} K\theta ^i .$$
Let $A:= p(\theta ),$ then 
$A,A\theta ,\ldots , A\theta ^{\ell -t-1 } $ 
are linearly independent over $K$. 
So the $k$-dimension of $A \End_k(K) $ is at least 
$\ell (\ell -t) $. Therefore the rank of $A$ is at least $\ell -t$
so the kernel of $A$ has at most dimension $t$ over $k$. 
\eb

\begin{cor} \label{dim}
If $v \in K^m$ has rank $r$ then
 $(v,\theta(v),\ldots , \theta ^{r-1}(v) )$ are linearly 
independent over $K$.
\end{cor}

\bew
Assume that there are $a_0,\ldots , a_{r-1} \in K$ such that 
$\sum _{i=0}^{r-1} a_i \theta ^i(v) = 0 $. 
Put $p:= \sum _{i=0}^{r-1} a_i x^i \in K[x] $. 
Then the kernel of $p(\theta )$ contains the  subspace 
$\langle v_1,\ldots v_m \rangle  _k \leq K$ of dimension $r$. 
As the degree of $p$ is $\leq r-1$  Lemma \ref{jordan} implies that 
$p=0$. 
\eb

\begin{cor}\label{gleich} 
Let $v\in K^m$ be a Gabidulin vector and $1\leq d \leq m-1$. 
Then 
$$\langle v \rangle _K = \theta ^{1-d}  ( \bigcap _{i=0}^{d-1} \theta ^i(\tilde{\cC }(v,\theta,d)) ) .$$
In particular 
 $\cC (v,\theta,d) = \cC (w,\theta ,d) $ if and only if $v=\alpha w$ for some 
$0\neq \alpha \in K$. 
\end{cor}

\bew
The inclusion $\subseteq  $ is clear. 
So let 
$x\in \bigcap _{i=0}^{d-1} \theta ^i(\tilde{\cC }(v,\theta,d)) $. 
By Corollary \ref{dim} the vectors 
$(v,\theta(v),\ldots , \theta ^{d} (v) ) $ and hence also 
$(\theta ^i(v), \theta ^{i+1}(v) ,\ldots , \theta ^{i+d}(v) ) $ are
linearly independent over $K$ (for all $i$). 
In particular there are unique $a_{ij} \in K$ such that 
$$x = \sum _{j=0}^{d-1} a_{ij} \theta ^{i+j} (v) \mbox{ for all } 0\leq i\leq d-1 .$$
$$ \begin{array}{lrccccccccc} \mbox{ So } &  x = & 
 a_{00} v & + & a_{01} \theta (v) & + &  \ldots & + & a_{0d-1} \theta ^{d-1} (v) & & \\ 
 & = &  &  & a_{10} \theta (v) & + &  \ldots & + & a_{1d-2} \theta ^{d-1} (v) & + & a_{1d-1} \theta ^d(v) \end{array} $$ 
which shows that $a_{00} = 0$, $a_{01} = a_{10} $, 
$\ldots $, $a_{0d-1} = a_{1d-2} $, $a_{1d-1} = 0$ because 
$(v,\theta(v),\ldots , \theta ^{d} (v) ) $ are linearly independent. 
Comparing the coefficients $a_{1j}$ and $a_{2j}$ we similarly find that 
$a_{10}  = 0 $, $a_{11} = a_{20}$, $\ldots $, $(0=) a_{1d-1} = a_{2 d-2} $,
 $a_{2d-1} = 0$. 
So recursively we find that $x= a_{0d-1} \theta^{d-1} (v)$. 
\eb

\begin{theorem}
Let $v=(v_1,\ldots , v_m) \in K^m$ be a Gabidulin vector. 
Then
$\dim _k (\cC (v,\theta ,d) )= \ell d  $ and 
 $d(\cC (v,\theta,d)) = m-d+1$. In particular 
Gabidulin-like codes are  \MRD codes. 
\end{theorem}

\bew
It follows from Corollary \ref{dim} that $\dim _K(\tilde{\cC} (v,\theta,d)) = d$. As $\epsilon _B$ is an isomorphism and $\dim _k(K) = \ell $, we get 
$\dim _k (\cC (v,\theta ,d) )= \ell d  $. 
To obtain the \MRD property it suffices to show that 
any non zero $C\in \cC (v,\theta,d) $ has rank $\geq m- (d-1) $.
Let
$$0\neq C= 
\epsilon _B (\sum _{i=0}^{d-1} a_i \theta ^i (v) ) \in \cC (v,\theta,d).$$
Then the right kernel of $C$ is 
$$ 
\{ b= (b_1,\ldots , b_m)^{tr} \in k^{m\times 1} \mid Cb = 0 \}  = 
\{ b \in k^{m\times 1} \mid \sum _{i=0}^{d-1} a_i \theta ^i ( \sum _{j=1}^m b_jv_j) = 0 \}  $$ 
and hence isomorphic to the 
 kernel of  the restriction
of $ \sum _{i=0}^{d-1} a_i \theta _i  $ to 
$\langle v_1,\ldots , v_m \rangle_k  $
because $(v_1,\ldots , v_m)$ are
linearly independent over $k$.
By Lemma \ref{jordan} 
the kernel of $\sum _{i=0}^{d-1} a_i \theta _i  \in \End_k(K)$ has 
dimension at most $d-1$, so also the right kernel of $C$ has dimension 
at most $d-1$ and hence the rank of $C$ is $\geq m - (d-1)$.
\eb

\begin{theorem} \label{cond}
A lifted \MRD code $\cC = \epsilon_B(\tilde{\cC }) \leq k^{\ell \times m}$ with
$\dim _K(\tilde{\cC }) = d < m$ is a Gabidulin-like code  if and only if 
$$\dim _K(\bigcap _{i=0}^{d-1} \theta ^i(\tilde{\cC }) )   = 1.$$
\end{theorem}

\bew
For Gabidulin-like codes the dimension of the intersection is 1
by Corollary \ref{gleich}. 
So it remains to show the converse direction: 
Assume that $\bigcap _{i=0}^{d-1} \theta ^i(\tilde{\cC })  =\langle x \rangle _{K}$. 
Then $x\in \theta ^{d-1}(\tilde{\cC})$, so there is a unique
$v= (v_1,\ldots , v_m) \in \tilde{\cC} $, such that 
$$x=\theta ^{d-1}(v) = \theta ^{d-2}(\theta (v) ) = \ldots = \theta (\theta ^{d-2}(v)) .$$ 
As $x =\theta ^{d-i-1} (\theta ^{i}(v)) \in \theta ^{d-i-1} (\tilde{\cC} )$ for 
all $0\leq i \leq d-1 $ the injectivity of $\theta $ implies that 
$\theta ^{i} (v) \in \tilde{\cC }$ for all $0\leq i \leq d-1 $. 
\\
We now show that $v$ has rank $m$. 
Assume that $\dim _k \langle v_1,\ldots , v_m \rangle = r < m$. 
Then there is some $h\in \GL_m(k)$ such that 
$(v_1,\ldots , v_m) h= (w_1,\ldots , w_r, 0,\ldots , 0)$. 
Clearly $d(\cC) = d(\cC h) $. 
Let $w:=(w_1,\ldots , w_r)$. Then $w$ is a Gabidulin vector of length $r$ 
and $\cC h $ contains $(\cD | 0^{\ell \times (m-r) })$, 
where $\cD $ 
is  the Gabidulin code 
$\cD = \cC (w,\theta ,d) \leq k^{\ell \times r}$ if $d\leq r-1$ and
$\cD  = k^{\ell \times r}$ if $d\geq r$. 
In the first case $d(\cD) =r-d+1 < m-d+1$ (because we assumed $r<m$) and 
$d(\cC) = 1 < m-d+1 $ (because $d<m$) in the second case.
This contradicts the assumption that $\cC $ is an \MRD code.
\\
So $v$ is a Gabidulin vector and hence the subcode 
$$\tilde{\cC} (v,\theta,d) = \langle v,\theta (v),\ldots ,\theta ^{d-1}(v) \rangle $$ of $\tilde{\cC }$ has dimension $d$, therefore 
$\tilde{\cC } = \tilde{\cC} (v,\theta,d) $.
\eb

\section{Computing automorphism groups  of rank metric codes.} \label{AUTO}

The $k$-linear rank distance preserving automorphisms of  $k^{\ell\times m} $ 
are the maps
$$\kappa _{g,h} : X \mapsto g^{-1} X h \mbox{ for } g\in \GL_\ell (k) , h\in \GL _m (k) $$
(see \cite[Theorem 3.4]{Wan}) and, if $\ell = m$, also 
$$X \mapsto g^{-1} X^{tr} h  \ \ ( g\in \GL_\ell (k) , h\in \GL _m (k) ) .$$
Two codes $\cC $ and $\cD $ in $k^{\ell\times m} $ are called 
(properly) equivalent, if $\cC = g^{-1} \cD h $ for some $g\in \GL_\ell (k) , h\in \GL _m (k)$
and 
$$\Aut (\cC ) := \{ (g,h) \in  \GL_\ell (k) \times \GL _m (k) \mid 
g^{-1} \cC h = \cC \} $$ is called the (proper) automorphism group of $\cC $. 
Note that $\kappa _{g,h} = \kappa _{ag,a^{-1}h }$ for all $0\neq a \in k$,
so that different automorphisms might induce the same mappings on 
$k^{\ell \times m}$. 

The following definition is fundamental in our algorithm to 
compute rank metric automorphism groups.

\begin{definition}\label{defideal}
Let $k$ be a field  and $\cC \leq k^{\ell \times m }$ a subspace of 
$\ell \times m $-matrices over $k$.
Then we define the right and left idealiser of $\cC$  as 
$$R(\cC ) = \{ Y\in k^{m\times m } \mid \cC Y \subseteq \cC \} \mbox{ and } 
L(\cC ) = \{ X\in k^{\ell \times \ell  } \mid X \cC  \subseteq \cC \}  .$$
\end{definition}

Then clearly $R(\cC )$ and $L(\cC ) $ are subalgebras of the full matrix algebra.
%Direct computations relate left and right idealisers of equivalent codes:
%
%\begin{rem}
%If 
%$\cC = g^{-1} \cD h $ for some $g\in \GL_\ell (k) , h\in \GL _m (k)$ 
%then $R(\cC ) = h^{-1} R(\cD ) h $ and 
%$L(\cC ) = g ^{-1} L(\cD ) g $. 
%\end{rem}

All lifted codes, in particular the Gabidulin-like codes from 
Definition \ref{Gabi}, 
 are invariant under left multiplication with
the field $K$, or more precisely its image under the regular 
representation $\Delta _B(K) \leq k^{\ell \times \ell}$, 
so $\Delta _B(K) \leq L(\cC ) $. 
Note that $K\cong \Delta _B(K) $ is a maximal subfield of the 
central-simple $k$-algebra $k^{\ell \times \ell}$.
The following lemma is probably well known but crucial, as it 
gives us all possible left idealisers of 
such $K$-linear codes $\cC $: 

\begin{lemma}\label{doublecentraliser} 
Let $K$ be a field extension of degree $\ell $ over $k$ and 
$B$ some $k$-basis of $K$. 
Let $A$ be a $k$-algebra with 
$$\Delta _B(K) \leq A \leq k^{\ell \times \ell } .$$
Then there is a subfield $k\leq F \leq K$ such that 
$$A=C_{k^{\ell \times \ell }}(\Delta _B(F)) \cong F^{s\times s}$$ with $s=[K:F]$.
\end{lemma} 

\bew
Let $A$ be a subalgebra of $k^{\ell \times \ell }$ containing $\Delta _B(K)$.
Then $k^{\ell \times 1}$ is a simple $A$-module, because it has no
$\Delta _B(K)$-invariant submodules. 
Also $k^{\ell \times 1}$ is a   faithful  $k^{\ell \times \ell} $-module
and hence also its annihilator 
in $A$ is trivial,
$\{ a \in A \mid a k^{\ell \times 1} = \{ 0 \} \} = \{ 0 \} $.
So $A$ has a faithful simple module and hence 
is a simple $k$-algebra. Therefore $A$ has the
double-centraliser property $A=C_{k^{\ell \times \ell }} (C)$ 
for $C:= C_{k^{\ell \times \ell }}  (A) $ (see \cite[Theorem 7.11]{Reiner}). 
Clearly 
$$k \subseteq C \subseteq C_{k^{\ell \times \ell }} (\Delta_B(K)) = \Delta _B(K) ,$$ so 
$C= \Delta _B(F)$ for some subfield $F$ of $K$.
\eb

\begin{cor}\label{left}
Let $K/k$ be an extension of degree $\ell $, 
$\cC = \epsilon_B(\tilde{\cC}) $ be a lifted code for some $K$-linear 
code $\tilde{\cC} \leq K^m $. Then 
there is a subfield $F$ with $k\leq F \leq K$ such that 
$$L(\cC )   = 
C_{k^{\ell \times \ell }}(\Delta _B(F)) \cong F^{s\times s}$$ with $s=[K:F]$.
\end{cor}

Let $R(\cC )^{\times } := R(\cC ) \cap \GL_m(k) $ and 
$L(\cC )^{\times } := L(\cC ) \cap \GL_{\ell }(k) $ denote the unit groups
of right and left idealiser. 
We also let 
$$
\begin{array}{ll} N(R(\cC )) :=  & \{ h \in \GL_m(k ) \mid h^{-1} R(\cC ) h = R(\cC) \} \mbox{ and } \\
N(L(\cC )) := & \{ g \in \GL_{\ell}(k ) \mid g^{-1} L(\cC ) g = L(\cC) \}. 
\end{array} $$
Clearly $R(\cC )^{\times } \leq N(R(\cC ))$ and 
 $L(\cC )^{\times } \leq N(L(\cC ))$. 
The algorithm described below only applies to rank metric codes
for which one of the indices is finite. 
Note that this is always the case if $k$ is a finite field, but also 
for all lifted codes.
In this case let
$$n(\cC ) := \gcd \{ [ N(R(\cC )) : R(\cC )^{\times } ], 
[ N(L(\cC )) : L(\cC )^{\times } ] \} $$ 
denote the greatest common divisor of these two indices, 
otherwise let $n(\cC ):= \infty $.
Let 
$$\begin{array}{lll} \pi _1: & \GL_{\ell} (k) \times \GL _m (k) \to \GL _{\ell }(k) , & (g,h ) \mapsto g  \\ 
 \pi _2: & \GL_{\ell} (k) \times \GL _m (k) \to \GL _{m }(k) , & (g,h ) \mapsto h  \end{array} $$ 
denote the projections onto the first and second component. 

\begin{theorem}\label{subdirect}
$$L(\cC )^{\times } \times R(\cC )^{\times }  \leq \Aut (\cC ) 
 \leq N(L(\cC )) \times N(R(\cC ))$$
The automorphism group 
$\Aut (\cC ) = \{  (g,h) \in N(L(\cC )) \times N(R(\cC )) \mid 
g^{-1} \cC h = \cC  \} $ 
satisfies $\pi _1(\Aut(\cC )) / L(\cC )^{\times }  \cong \pi _2 (\Aut(\cC ))/R(\cC )^{\times }$. 
In particular the order of the factor group
$\Aut(\cC ) / (L(\cC )^{\times } \times R(\cC )^{\times } )  $ 
divides $n(\cC )$.
\end{theorem}

\bew
The first two statements are clear, we only need to 
prove the isomorphism (which is also a standard argument): 
By abuse of notation we denote by $\pi _i$ the restriction of
$\pi _i$ to $G:=\Aut (\cC )$ and put $L:= L(\cC)^{\times } $ and 
$R:= R(\cC)^{\times }$. 
Then
$$L \cong \{ (g,1) \mid g\in L  \} = \ker (\pi _2) \mbox{ and } 
R \cong \{ (1,h) \mid h\in R  \} = \ker (\pi _1) .$$
Define the two group epimorphisms
 $$\overline{\pi _1} :G \to \pi _1(G)/ L , (g,h) \mapsto g\cdot L
\mbox{ and } 
\overline{\pi _2} :G \to \pi _2(G)/ R , (g,h) \mapsto h\cdot R .$$
Then $\ker (\overline{\pi _1} ) = \ker (\overline{\pi _2}) = L\times R $ 
and hence  
$$ N(L(\cC)) / L \geq  \pi _1(G)/ L  \cong G/(R\times L)  \cong 
\pi_2(G)/R \leq N(R(\cC ))/R .$$
In particular $G/(R\times L)$ is isomorphic to a subgroup of 
$N(L(\cC))/L$ and $N(R(\cC ))/ R$, therefore its order divides
the order of both factor groups.
\eb

To compute $\Aut (\cC ) $ we first compute $L(\cC )$ and $R(\cC )$
as the intersection of two subspaces.
More general for $\cC , \cD \leq k^{\ell \times m} $ put 
$$ L(\cC ,\cD ) := \{ X \in k^{\ell \times \ell } \mid X \cC \subseteq \cD \} 
\mbox{ and } 
 R(\cC ,\cD ) := \{ Y \in k^{m \times m } \mid  \cC Y \subseteq \cD \} .$$

We have 
$$k^{\ell \times m} \cong k^{\ell} \otimes k^{m} \cong k^{\ell m} .$$
The linear mappings of this $\ell m$-dimensional vector space induced 
by left multiplication by elements in $k^{\ell \times \ell}$ 
form the subalgebra 
$$ A:= k^{\ell \times \ell }\otimes k \leq 
k^{\ell \times \ell} \otimes k^{m\times m} \cong k^{\ell m \times \ell m}  $$ 
and similarly those induced by right multiplication 
$$ B:= k\otimes k^{m\times m } \leq 
k^{\ell \times \ell} \otimes k^{m\times m} \cong k^{\ell m \times \ell m}  .$$ 
For 
$\cC,\cD \leq k^{\ell \times m} \cong k^{\ell m}$  
let $LR(\cC,\cD ) := \{ X \in k^{\ell m \times \ell m} \mid \cC X \subseteq \cD \} $.
Then  bases of 
$$L(\cC, \cD ) = A \cap LR(\cC, \cD ) \mbox{ and } R(\cC, \cD ) = B\cap LR(\cC, \cD ) $$
can be computed using Zassenhaus' algorithm for computing 
intersections of subspaces.

In general normalizers of subalgebras are hard to compute. 
However, at least for finite fields, there are fast algorithms to
compute the normaliser of a subgroup of the general linear group
\cite{Colva}. 
Clearly $N(L(\cC )) \leq N_{\GL_{\ell }(k)} (L(\cC )^{\times}) $ 
with equality if $L(\cC ) $
is generated by its unit group.
The same holds for $N(R(\cC) )$. 

For lifted codes, we always have 
$L(\cC) \cong F^{s \times s}$ for some $k\leq F \leq K$ and hence 
 $$N(L(\cC)) = N_{\GL_{\ell }(k)} (L(\cC)^{\times}) \cong \GL_s(F)  . \Gal(F/k) .$$
From now on we assume that we know one of 
$N(L(\cC ))$ and $N(R(\cC)) $. 
If both are known, then we choose the one $(X=L, R)$ for which 
the index $[N(X(\cC)) : X(\cC) ^{\times} ]$ is smaller. 
To ease notation assume that $X=L$. 
Let 
$$N(L(\cC) ) = \disj _{j=1}^n t_j L(\cC )^{\times } .$$
Put $J:=\{ \}$. 
For every $j=1,\ldots ,n$ we
compute $R(t_j\cC ,\cC) $ as described above. 
If this space  contains an invertible matrix 
 $s_j $ then
 put $J:= J \cup \{ (t_j,s_j) \} $. 

Now Theorem \ref{subdirect} implies that we obtain
a generating set of the automorphism group as follows.

\begin{theorem}
Let $R$ respectively $L$ generating sets of $R(\cC )^{\times }$ 
respectively $L(\cC )^{\times }$ and $J, s_j,t_j$ be as above. Then
$$\Aut (\cC) = \langle (g,1) , (1,h) , (t_j,s_j) \mid g\in L, h \in R, 
(t_j,s_j) \in J \rangle .$$
\end{theorem} 

A similar strategy can be used to compute equivalences between 
rank metric codes.

\section{Automorphism groups of Gabidulin-like codes} \label{GABI}

In the whole section we assume that 
 $K/k$ is a cyclic extension of degree $\ell $  and choose 
a generator $\theta $ of the Galois group $\Gal (K/k)$. 
For $k$-linearly independent $v:= (v_1,\ldots , v_m) \in K^m$  and
$0\leq d < m$ 
the Gabidulin-like code 
$\cC (v,\theta,d) := \epsilon _B (\langle v,\theta(v),\ldots , \theta ^{d-1}(v) \rangle _K )$ is defined in Definition \ref{Gabi}. 

To compute 
 the right idealiser (cf. Definition \ref{defideal}) of a Gabidulin-like code
$\cC (v,\theta , d)$ 
we define 
$$V_v := \langle v_1,\ldots , v_m \rangle _{k} \leq K $$ 
be the $k$-subspace of $K$ generated by the entries of the Gabidulin 
vector $v=(v_1,\ldots , v_m)$. This is an $m$-dimensional subspace of 
$K$. 

\begin{lemma} (\cite[IV.4]{Morrison} for finite fields) 
Let $k\leq M \leq K$ be the maximal subfield of $K$ such that $V_v$ 
is an $M$-linear subspace of $K$. 
Then $R(\cC (v,\theta,d) ) \cong M $ for all $0< d < m $. 
\end{lemma} 

\bew
Let $0 \neq Y \in k^{m\times m }$ such that 
$\cC (v,\theta , d) Y \subseteq \cC (v,\theta ,d) $.
Then by Corollary \ref{gleich} 
$$\langle v Y\rangle _K  = \theta ^{1-d}  ( \bigcap _{i=0}^{d-1} \theta ^i(\tilde{\cC }(v,\theta,d) Y ) ) \subseteq \langle v \rangle _K $$
so there is some $\alpha \in K$ such that 
$vY = \alpha v$. 
Moreover $v Y \in V_v $ because the entries of $Y$ are in $k$.
So $\alpha \in M $. 
\eb

 To compute the left idealiser we introduce the splitting field of a Gabidulin-like code.

\begin{definition}
Let $\tilde{\cC} \leq K^m$ be a Gabidulin-like code. The smallest subfield $k \leq F \leq K$ such that there exists a subspace $\tilde{\cD} \leq F^m$ satisfying 
$$\tilde{\cC} = \tilde{\cD} \otimes_F K$$
is called the splitting field of $\tilde{\cC}$.
\end{definition}

\begin{lemma}
Let $\tilde{\cC} \leq K^m$ be a Gabidulin-like code 
with splitting field $F$ and let
$\tilde{\cD} \leq F^m$ with $\tilde{\cC} = \tilde{\cD} \otimes_F K$. 
Then $\tilde{\cD}$ is also a Gabidulin-like code.
\end{lemma}

\bew Let $x \in \tilde{\cD}$. Then the rank of $x \in F^m$ 
equals the rank of $x\otimes 1 \in (F\otimes _F K)^m $. 
As $\tilde{\cC}$ is a MRD code, so is $\tilde{\cD}$. 
Now let $d = \dim_K(\tilde{\cC}) = \dim_F(\tilde{\cD})$. 
As the intersection of vector spaces commutes with the tensor product and $K$ is fixed (as a set) by all powers of $\theta$, we get
$$\bigcap_{i=0}^{d-1} \theta^i(\tilde{\cC}) = \left(\bigcap_{i=0}^{d-1} \theta^i(\tilde{\cD})\right) \otimes_F K.$$ Applying Theorem \ref{cond} the intersection on the left hand side has $K-$dimension $1$. Thus the intersection
$\bigcap_{i=0}^{d-1} \theta^i(\tilde{\cD})$
  has $F-$dimension $1$ and as $\tilde{\cD}$ is an MRD code, 
Theorem \ref{cond} implies that $\tilde{\cD}$ is a Gabidulin-like code. \eb

The next lemma 
 allows us to compute the splitting field using only the Gabidulin vector. Note that the extension $F/k$ is also cyclic and 
$\Gal (F/k)$ is generated by $\theta_{|F}$.

\begin{proposition}
Let $\tilde{\cC} \leq K^m$ be a Gabidulin-like code with splitting field $F$ and Gabidulin vector $v$ normalised so that $v_1 = 1$. Then $F = k[v_2,\ldots, v_m]$.
\end{proposition}

\bew Let $F' := k[v_2,\ldots, v_m]$, $F$ the splitting field of $\tilde{\cC}$ and $d := \dim_K(\tilde{\cC})$. 
Let $w \in F^m$ be the normalised Gabidulin vector of $\tilde{\cD} \leq F^m$
 and set $w' := w \otimes_F 1 \in \tilde{\cC}$. 
Then $w'$ has rank $m$ and 
$\theta^i(w') \in \tilde{\cC}$ for all $0 \leq i \leq d-1$.
So $w'$ is a Gabidulin vector for $\tilde{\cC}$ and thus by Corollary \ref{gleich} a multiple of $v$. 
As both vectors are normalised we get $v = w'$, which gives us $F' \leq F$. \\
For the other direction define 
$\tilde{\cD} := \tilde{\cC}(v,\theta_{|F'},d) \leq (F')^m$ by interpreting $v$ as an element of $(F')^m$. Then $\tilde{\cC} = \tilde{\cD} \otimes_{F'} K$ and the minimality of the splitting field gives us $F \leq F'$. \eb

If we take a Gabidulin-like code $\tilde{\cC} = \tilde{\cD} \otimes_F K$ with splitting field $F$ and a basis $B$ adjusted to the decomposition $K = F \otimes_F K$, we get
$$\cC = \cD^{s \times 1}$$
where $s = [K:F].$ This allows us to compute the left idealiser.

\begin{theorem}
Let $\cC (v,\theta,d) $ be a Gabidulin-like code with splitting field $F \leq K$. Then
$$L(\cC (v,\theta,d)) = C_{k^{\ell \times \ell }} (\Delta_B(F)) \cong F^{s\times s}$$ (with $s = [K:F]$).
\end{theorem}
\bew We can change $B$ to fit the decomposition $K = F \otimes_F K$ as mentioned above. Then $\cD^{s \times 1}$ is a $F^{s \times s}$-module. For the other direction of the equality we use Corollary \ref{left}. As we always have $\Delta_B(K) \leq L(\cC(v,\theta,d))$, there is some field $F'$ such that $L(\cC(v,\theta,d)) \cong (F')^{s' \times s'}$ where $s' = [K:F'].$ Then $\cC$ is equivalent to $(\cD')^{s' \times 1}$ for some $\cD'$ as these are the only $(F')^{s' \times s'}$-modules. The minimality of the splitting field now gives us $F = F'$\eb

Putting together all the results of this section, we now obtain 
the following structure of the automorphism group of 
Gabidulin-like codes:

\begin{theorem} 
Let $v=(v_1,\ldots , v_m) \in K^{m} $ be a Gabidulin vector normalised so that 
$v_1 = 1$. Let $k\leq M \leq K $ be the maximal subfield of $K$ such that
$$V_v := \langle v_1,\ldots , v_m \rangle _{k} \leq K $$ 
is an $M$-linear subspace of $K$.
Let $F= k[v_2,\ldots , v_m ]$ be the minimal subfield of $K$ that contains $V_v$
and $s:=[K:F]$. 
Then there is a subgroup $G\leq \Gal (F/k) =\langle \theta _{|F} \rangle $
such that for any $1\leq d < m$  
$$\Aut (\cC (v,\theta,d)) \cong (\GL_s(F) \times M^{\times }) . G .$$
\end{theorem}

\end{document}